\documentclass[floatfix, reprint, aps, pra, amsmath, amssymb]{revtex4-2}

\usepackage{graphicx}% Include figure files
\usepackage{dcolumn}% Align table columns on decimal point
\usepackage{bm}% bold math
\usepackage{verbatim}
\usepackage{braket}
\usepackage{enumitem} % Allow letters for list enumeration
\usepackage{soul}
\usepackage{textcomp, gensymb} % Use degree symbol in math enviroment
\usepackage{hyperref}% add hypertext capabilities
\usepackage{makecell}
\usepackage{xcolor}
\usepackage[font=small]{caption}
\usepackage[labelformat=empty, position=top]{subcaption}
\usepackage[export]{adjustbox}

\begin{document}

\preprint{APS/123-QED}

\title{Milestone toward an  ECRIPAC accelerator demonstrator}% Force line breaks with \\

\author{Andrea Cernuschi}
 \email{andrea.cernuschi@lpsc.in2p3.fr}
\author{Thomas Thuillier}%
 \email{thomas.thuillier@lpsc.in2p3.fr}
\affiliation{%
 Université Grenoble Alpes, CNRS, Grenoble INP, LPSC-IN2P3, 38000 Grenoble, France
}%
\author{Laurent Garrigues}
\email{laurent.garrigues@laplace.univ-tlse.fr}
\affiliation{Université de Toulouse, Toulouse INP, CNRS, LAPLACE, Toulouse, France}

% ****** SECTION BREAK ******

\begin{abstract}
The Electron Cyclotron Resonance Ion Plasma ACcelerator (ECRIPAC) is an original accelerator concept proposed in the nineties for the generation of highly energetic pulsed ion beams, suitable for a wide array of applications.
The initial studies on the subject were characterized by an important calculation mistake, leading to an incomplete and erroneous literature on the topic. Nevertheless, the simple and well mastered techniques involved in the system (radio frequency and magnetic field), together with the device compactness, are strong motivations for further studies on ECRIPAC.
This work proposes a comprehensive introduction to the ECRIPAC accelerator physics, including a summary of its corrected theory.
The designs of several compact demonstrator devices, able to accelerate different ion species to energies up to 100 MeV, are presented.
A particular focus is devoted to a He\textsuperscript{2+} accelerator, capable of generating 9.5 MeV/nucleon ions inside a 1.8 m long accelerating cavity. This device has been simulated using a Monte-Carlo (MC) code, developed to model the electron dynamics inside this system. The MC results show an excellent agreement with the updated theory, which validates the new theoretical framework of ECRIPAC.
Finally, some estimations for the beam parameters of the ion bunch extracted from the accelerator are provided.
\end{abstract}

\maketitle

% ****** SECTION BREAK ******
\textit{Introduction --}
The Electron Cyclotron Resonance Ion Plasma ACcelerator (ECRIPAC)~\cite{Geller-ecripac}, proposed by R.~Geller and K.~Golovanivsky, is an original accelerator concept for the generation of pulsed ion beams with adjustable energy.
Unlike conventional accelerator techniques, such as the typical linear accelerator (Linac) structures and their associated injectors~\cite{Ostroumov-linac,Stockli-injector,Zhang-injector}, plasma-based accelerators are often distinguished by their compact size. This is true for both well-established methods like plasma wakefield acceleration~\cite{Gorbunov-prl, Hogan-prl} and more original concepts such as autoresonant accelerators~\cite{Sloan-prl} and ECRIPAC.
An advantage of ECRIPAC is that it requires simple, robust, and well-mastered technologies for its operation, leading to possible application in many fields, especially in the medical domain~\cite{Schwartz-ecripac,Ishibashi-ecripac,Inoue-ecripac}.
Due to an important calculation mistake reported in the initial paper~\cite{Geller-ecripac}, the most recent literature on ECRIPAC~\cite{Ishibashi-ecripac,Inoue-ecripac} is based on incomplete and possibly wrong premises. Hence, a careful review of the physical theory underlying the ECRIPAC working principle has been carried out and it is presented in a specific paper~\cite{Cernuschi-ecripac}.
ECRIPAC exploits two known and experimentally verified physical principles: gyromagnetic autoresonance (GA)~\cite{Golovanivsky-autoresonant_acceleration,Golovanivsky-gyromagnetic_autoresonance,Golovanivsky-gyrac} and ion entrainment~\cite{Consoli-plasma_acceleration,Bardet-ion_entrainment, Bardet-pleiade}.
\\
GA~\cite{Golovanivsky-autoresonant_acceleration,Golovanivsky-gyromagnetic_autoresonance,Golovanivsky-gyrac} is a physical mechanism for heating electrons confined in a time-increasing magnetic field $B(t)$ by interaction with an injected microwave radiation (HF), whose frequency $\omega_{HF}$ satisfies the electron cyclotron resonance (ECR) condition for cold electrons

\begin{equation}
\Omega(t_0)=\frac{eB(t_0)}{m\gamma(t_0)}=\frac{eB_0}{m}=\omega_{HF} \; ,
\end{equation}

where $\Omega$ is the electron gyrofrequency, $m$ is the electron rest mass and $\gamma$ the Lorentz factor.
If the magnetic field satisfies some conditions related to a sufficiently smooth time behavior, the electron energy increases quasi-synchronously with the magnetic field growth in a so-called gyromagnetic autoresonance regime, overcoming the relativistic limitations of ECR:
\begin{equation}
    \gamma (t) \approx \frac{B(t)}{B_{0}}\;.
    \label{eq:gyrac_autoresonance}
\end{equation}
\\
Ion entrainment~\cite{Consoli-plasma_acceleration,Bardet-ion_entrainment, Bardet-pleiade} is the phenomenon responsible for ion acceleration, exploiting the energetic electrons generated by GA. The electron population of the plasma is longitudinally displaced from the ions using a magnetic field gradient, generating a space-charge field due to the difference in the local densities of the two plasma populations. This field accelerates the ions in the direction of the force acting on the electrons, without the need of any external electric fields.
The axial velocities of the electrons and ions have been experimentally observed to be quite similar~\cite{Bardet-ion_entrainment}, resulting in a net conversion of perpendicular electron energy $W_e$ to parallel ion energy $W_i$

\begin{equation}
    W_{e\perp}^{in}-W_{e\perp}^{fin}\approx W_{i\parallel}^{fin}\;,
    \label{eq:pleiade}
\end{equation}
where the parallel ($\parallel$) and perpendicular ($\perp$) energy orientations are expressed with respect to the magnetic field axis, and the indexes $in$ and $fin$ stand for initial and final respectively.
The overall ion acceleration uses well-established and commercially available technologies, consisting of coils to generate the overall magnetic field structure and a microwave injector, highlighting the interest of ECRIPAC for the particle accelerator community.
In this work, the ECRIPAC accelerator concept is first presented, along with a summary of the key points of the recently revised theory.
Next, we present the theoretical designs of several compact devices for the acceleration of ions relevant to medical applications, including a demonstrator for $He^{2+}$ ions up to approximately 10 MeV/A. The theoretical electron behavior inside the latter is then compared with the results of a Monte-Carlo simulation, developed to validate the electron dynamics in the machine. The same design is also used to present some estimation for the beam parameters of the ion bunches extracted from the accelerator.
Finally, conclusions and prospects are drawn.

% ****** SECTION BREAK ******
\begin{figure}[b]
    \centering
    \includegraphics[width=1.\linewidth, valign=t]{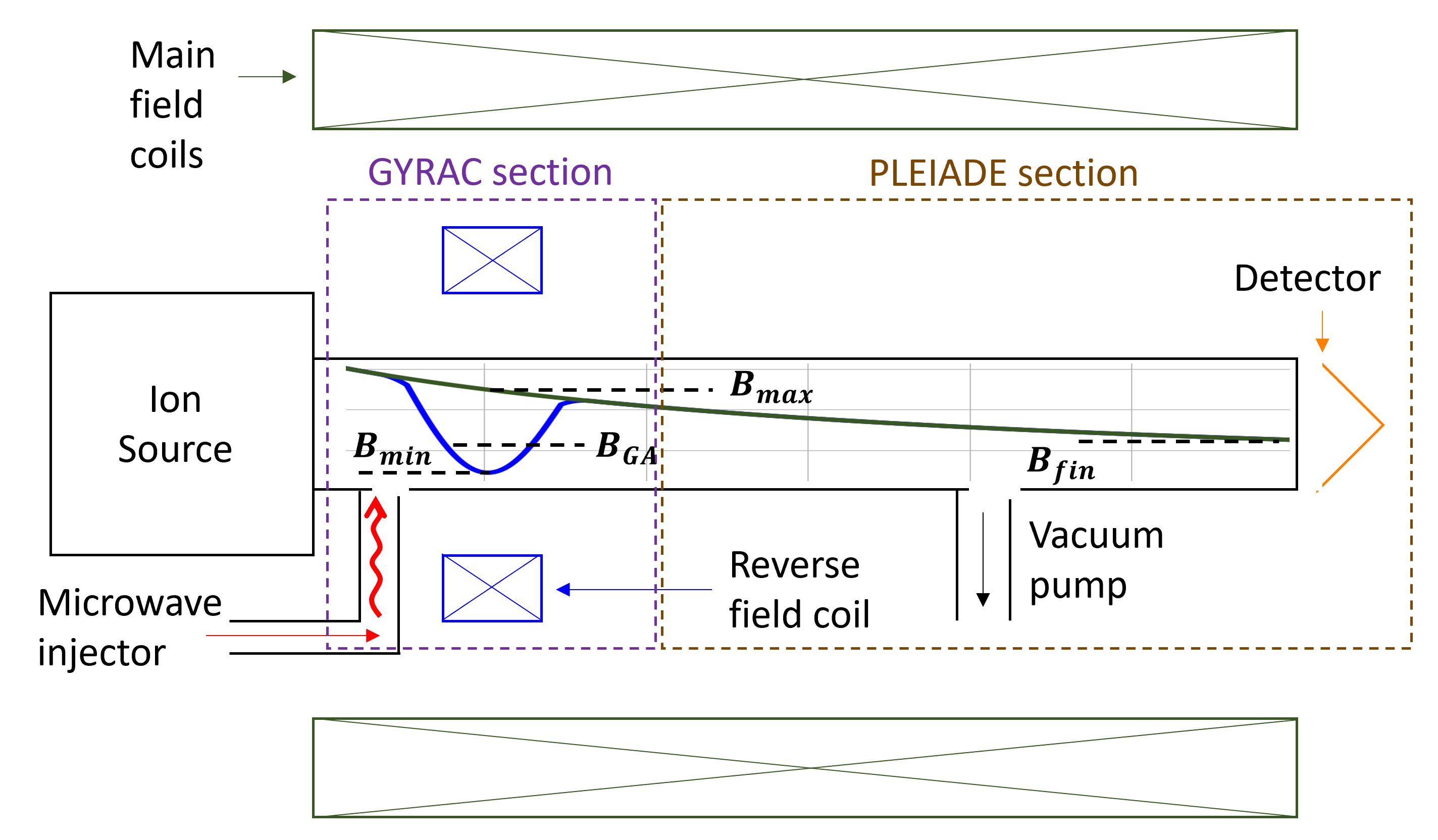}
    \caption{Schematic representation of ECRIPAC and its magnetic field. $B_{min}$, $B_{GA}$ and $B_{max}$ are the magnetic field strengths at the center of the magnetic mirror at the beginning, end of GA phase and end of the transient field behavior respectively, and $B_{fin}$ is the final magnetic field value at the end of the PLEIADE cavity.}
    \label{fig:ecripac}
\end{figure}

\textit{ECRIPAC working principle --}
An ECRIPAC device~\cite{Geller-ecripac} is composed of three sections (see Fig.~\ref{fig:ecripac}):

\begin{itemize}
    \item An injector of low pressure highly ionized plasma in free diffusion through differential pumping. A possible candidate for this stage is an ECR ion source~\cite{Geller_book}.
    \item A GYRAC section, which includes a reverse field pulsed coil, a resonant cavity and a microwave injection port.
    \item A PLEIADE section, constituted by a beam cavity and an axial distribution of coils, generating a long axial magnetic gradient encompassing the GYRAC section.
\end{itemize}

A typical ECRIPAC axial magnetic field profile is also reported in Fig.~\ref{fig:ecripac} inside the cavity: the main static magnetic field gradient (PLEIADE section) is displayed in dark green, with the maximum effect of the reversed pulsed magnetic field (GYRAC section) superimposed in blue.
The ECRIPAC working cycle is composed of three successive phases: the gyromagnetic autoresonance (GA) phase, the plasma compression (PC) phase and the PLEIADE (PL) phase.
\\
The ECRIPAC process~\cite{Geller-ecripac} starts with the GA phase, by triggering the reverse pulsed coil and hence delivering an axial sinusoidal magnetic field.
When the pulsed coil reaches its maximum intensity (time $t=0$), resulting in a minimum field $B_{min}\leq B_{0}$, both the plasma (coming from the injector) and the HF power are injected into the cavity. The plasma is then trapped inside the generated transient magnetic mirror, and the electrons are immediately heated by the ECR mechanism.
If the reverse pulsed magnetic field temporal evolution respects the GA hypothesis for a sufficiently slow-varying field in time~\cite{Golovanivsky-gyromagnetic_autoresonance}, the energy of the electrons increases according to Eq.~\ref{eq:gyrac_autoresonance}. 
\\
The plasma compression phase~\cite{Geller-ecripac} starts as soon as the HF wave injection is stopped ($t=t_{GA}$), and lasts until the pulsed magnetic mirror well vanishes ($t=t_{pul}$), corresponding to the complete restoration of the initial main magnetic field (dark green line in Fig.~\ref{fig:ecripac}).
During this period, the energy of the electron population increases by means of the electric field induced by the time variation of the magnetic field. However, the most prominent phenomenon taking place during this phase is the compression of the plasma, leading to increased electron and ion densities, which are crucial factors for the following ion acceleration.
Indeed, under the hypothesis of slowly varying magnetic field, it is possible to demonstrate that the electron motion inside ECRIPAC during this phase is characterized by the adiabatic constants of motion

\begin{equation}
    p_e^2/B=const \hspace{.5 cm} \text{and} \hspace{.5 cm} r_{e}^2B=const \; ,
    \label{eq:eq_motion}
\end{equation}

where $p_e$ and $r_e$ are respectively the electron momentum and radial distance from the cavity axis.
Thus, the increasing magnetic field over time leads to a decrease of $r_{e}$, radially compressing the plasma. Moreover, the axial oscillatory motion of the electrons inside the magnetic mirror is damped in time, leading also to an axial compression of the plasma disk.
Starting from the equation of motion for the electrons, it is possible to obtain the time evolution of the Lorentz factor during the plasma compression phase ($t_{GA}\leq t \leq t_{pul}$):
\begin{equation}
    \gamma(t)=\sqrt{1+(\gamma_{GA}^2-1)\frac{B(t)}{B_{GA}}} \;,
    \label{eq:ecripac_compression_gamma}
\end{equation}

where $\gamma_{GA}=\gamma(t_{GA})$.
\\
The PLEIADE phase~\cite{Geller-ecripac}, starting at $t=t_{pul}$, is the last part of the ECRIPAC working cycle. Its objective is the acceleration of the ion population.
Due to the adiabatic conservation of the magnetic dipole moment, the negative gradient of the magnetic field converts the electron perpendicular velocity into parallel velocity, which leads to the charge separation inside the plasma and the consequential ion acceleration through the ion entrainment mechanism.
An important condition that the PLEIADE magnetic field must satisfy is $d^2B/dz^2>0$, in order to avoid possible macroscopic instabilities which could prevent the ion acceleration.
Luckily, the diamagnetic character of the plasma helps to mitigate this condition~\cite{Geller-design_ecripac}.
\\
The accelerator stability during the PLEIADE phase actually imposes more stringent conditions on the magnetic field profile, being constrained by two main phenomena: the stability of the electron bunch and the ion shake-out (some ions are not accelerated by the electrons and detach from the plasma).
An ion with mass $A m_a$ and electric charge $Ze$ (where $m_a$ is the atomic mass unit, $e$ the elementary charge, $A$ and $Z$ the mass and charge numbers respectively) is entrained by the electrons when the Coulomb attraction overcomes the electrons' acceleration (non shake-out condition), resulting in the following expression depending on the PLEIADE field profile

\begin{equation}
    \left|\frac{\nabla B_z}{B_z}\right|\leq \frac{2e}{m_{a}c^2}\frac{Z}{A}E_{sc}\;,
    \label{eq:ecripac_nsc}
\end{equation}

where $E_{sc}$ is the space-charge electric field generated inside the plasma.
Any ion with the ratio $A/Z$ not respecting Eq.~\ref{eq:ecripac_nsc} is hence shaken out from the plasma.
If $\left|\frac{\nabla B_z}{B_z}\right|> \frac{2e}{m_{a}c^2}E_{sc}$, Eq.~\ref{eq:ecripac_nsc} is never satisfied (since the lowest possible value of $A/Z$ is 1 considering protons, which corresponds to the highest possible value of the right-hand side of Eq.~\ref{eq:ecripac_nsc}).
This allows to use ECRIPAC as an electron accelerator using very large magnetic field gradients.
The stability of the electron bunch during the PLEIADE stage depends on both the spatial divergence of the electron bunch due to Coulomb repulsion and its dampening due to relativistic effects. Thus, it is possible to obtain an expression by comparing the radial and axial forces acting on the electrons

\begin{equation}
    \left|\frac{\nabla B_z}{B_z}\right|\geq \frac{2e}{m_ec^2(\gamma^3-\gamma)}E_{sc}\;.
    \label{eq:ecripac_esc}
\end{equation}

The minimum electron energy to accelerate a given ion can be obtained by equating Eqs.~\ref{eq:ecripac_nsc} and \ref{eq:ecripac_esc}

\begin{equation}
    \gamma_{min}^3-\gamma_{min}=\frac{m_{a}}{m_e}\frac{A}{Z} \;.
    \label{eq:ecripac_gamma_min}
\end{equation}

Finally, considering a number of ions much lower than electrons ($N_i \ll N_e$), the energy of the ions per nucleon at the end of the PLEIADE phase can be estimated as

\begin{equation}
    \frac{W_i}{A}\approx\frac{1}{2}\frac{\gamma^2_{PC}-1}{\gamma^2_{PC}}m_{a} c^2\left(1-\frac{B_{fin}}{B_{max}}\right) \;,
    \label{eq:ecripac-ion_energy_simple}
\end{equation}
where \textit{c} is the speed of light and $\gamma_{PC}=\gamma(t_{pul})$.

% ****** SECTION BREAK ******
\begin{table*}[t!]
\caption{\label{tab:prototype}
Parameters design for several ECRIPAC compact devices, able to accelerate ions of interest for the medical domain, including protons at 10 and 100 MeV and He$^{2+}$ and C$^{4+}$ ions up to 10 MeV/nucleons. The common parameters for all the designs are the heating frequency of microwave $f_{HF}=2.45$ GHz, the maximum magnetic field $B_{max}=5$ T, a conservative (low) initial electron density $n_e=15\%\;n_{cr}=1.12\cdot 10^{10}$ cm$^{-3}$ ($n_{cr}=(\epsilon_0m_e\omega_{HF}^2)/e^2$) and the rise time for the pulsed field $t_{pul}=50$ $\mu$s.
}
\begin{ruledtabular}
\begin{tabular}{lcccc}
Parameter & p$^+$ & p$^+$ high W & He$^{2+}$ & C$^{4+}$\\
\colrule
Final value of PLEIADE magnetic field ($B_{fin}$ in T) & 4.89 & 3.91 & 4.89 & 4.87\\
PLEIADE cavity length ($l_{PL}$ in m) & 0.6 & 9.0 & 1.8 & 3.9\\
Electron energy at the end of plasma compression phase ($W_{e,PC}$ in MeV) & 6.1 & 6.0 & 7.8 & 8.8\\
Estimated final ion energy per nucleon ($W_i/A$ in MeV/nucleon) & 10.01 & 99.88 & 9.53 & 10.04\\
\end{tabular}
\end{ruledtabular}
\end{table*}

\textit{Prototype designs for compact demonstrator devices --}
The theoretical study presented in \cite{Cernuschi-ecripac} shows more stringent limitations on the accelerator design than the ones previously presented. Notably, the minimum electron energy necessary at the beginning of the PLEAIDE phase is found to be higher than the formerly reported limit (Eq.~\ref{eq:ecripac_gamma_min}), due to the evolution of the electron bunch radius and energy. Moreover, the accelerator stability depends on several physical parameters. Overall, ECRIPAC is better suited to accelerate low A/Z ions using specifically tuned microwave and coil settings. A denser plasma is beneficial for the accelerator stability but can also be a source of plasma instabilities. 
Using these results, we propose in Table~\ref{tab:prototype} different designs to accelerate several ions of interest for the medical domain~\cite{Kraft-ions}, including protons at 10 and 100 MeV and He$^{2+}$ and C$^{4+}$ ions up to 10 MeV/nucleons.
The common parameters for all the designs are the heating frequency of microwave $f_{HF}=2.45$ GHz, the maximum magnetic field $B_{max}=5$ T, a conservative (low) initial electron density $n_e=1.12\cdot 10^{10}$ cm$^{-3}$ and the rise time for the pulsed field $t_{pul}=50$ $\mu$s.
It is worth noting that any other ion species with the same A/Z ratio can be accelerated in the same machine.
\\
The considered heating frequency for all the designs (2.45 GHz) is very common in industry.
The rise time indicated for the pulsed magnetic field, fixed at 50 $\mu$s, has been chosen to reduce the computational cost of the simulations presented in the next section, despite being an unrealistic value for a pulsed coil generating a magnetic field close to 5 T, even at very low duty cycles. This choice is justified as follows. First, the magnetic field variation over time should not theoretically influence the gyromagnetic autoresonance process, except for extremely large values ($dB/dt<E_{HF}\cdot\omega_{HF}$ using Gaussian units~\cite{Golovanivsky-gyrac, Golovanivsky-gyromagnetic_autoresonance}), well below the value of interest for the proposed design ($dB/dt\approx0.98$ G/ns for a 50 $\mu$s rise time and the proposed design, compared to a maximum value of $ E_{HF}\cdot\omega_{HF}\approx~25.63$~G/ns for the lowest HF electric field value tested $E_{HF}=0.5$ kV/cm).
This behavior has also been verified through numerical Monte Carlo simulations (see next section).
Moreover, several numerical simulations of GYRAC devices~\cite{Andreev-gyrac2012,Andreev-gyrac2016,Andreev-gyrac2017, Andreev-longgyrac2017, Andreev-longgyrac2020, Andreev-longgyrac2021}, which exploit gyromagnetic autoresonance as their working principle, has been successfully compared against experimental results, despite having artificially decreased the rise time of the pulsed field by a factor of 200-300 with respect to the experimental setup.
Hence, the rise time for a possible experimental prototype can be much longer to facilitate the technological challenges of the accelerator, while keeping all the other parameters fixed.
The electron plasma density has been fixed at 15\% of the critical plasma density ($n_{cr}=(\epsilon_0m_e\omega_{HF}^2)/e^2$) for the considered heating frequency of the microwave, resulting in a low pressure plasma. This choice has been made to avoid problems related to working with a denser plasma, such as complicated wave-plasma interactions.
\\
In the rest of the letter, a particular focus will be devoted to the He$^{2+}$ compact accelerator demonstrator, able to accelerate ion bunches to estimated energies on the order of $\sim 9.5$ MeV/A in a 1.8 m long PLEIADE section.
The same parameters can also be used to accelerate electrons at 8 MeV by using a steeper magnetic field gradient along the PLEIADE cavity (reducing $B_{fin}$).
As a comparison, a cyclotron capable of producing such an ion beam would require a magnetic pole diameter of approximately 3 m and an external multi-charged ion source, coupled to a 4 m long low-energy beam transport (LEBT) line to inject the beam into the cyclotron.
Otherwise, a linear accelerator would require at least an ion source with a 4 m long LEBT, a 7 m long radio-frequency quadrupole to pre-accelerate the ions up to 1 MeV/A, a 4 m long medium-energy beam transport (MEBT) line, followed by a set of 1 m long RF cavities to reach 9.5 MeV/A.

% ****** SECTION BREAK ******
\begin{figure}[t]
\begin{subfigure}{1\linewidth}
  \centering
  \includegraphics[width=1\linewidth]{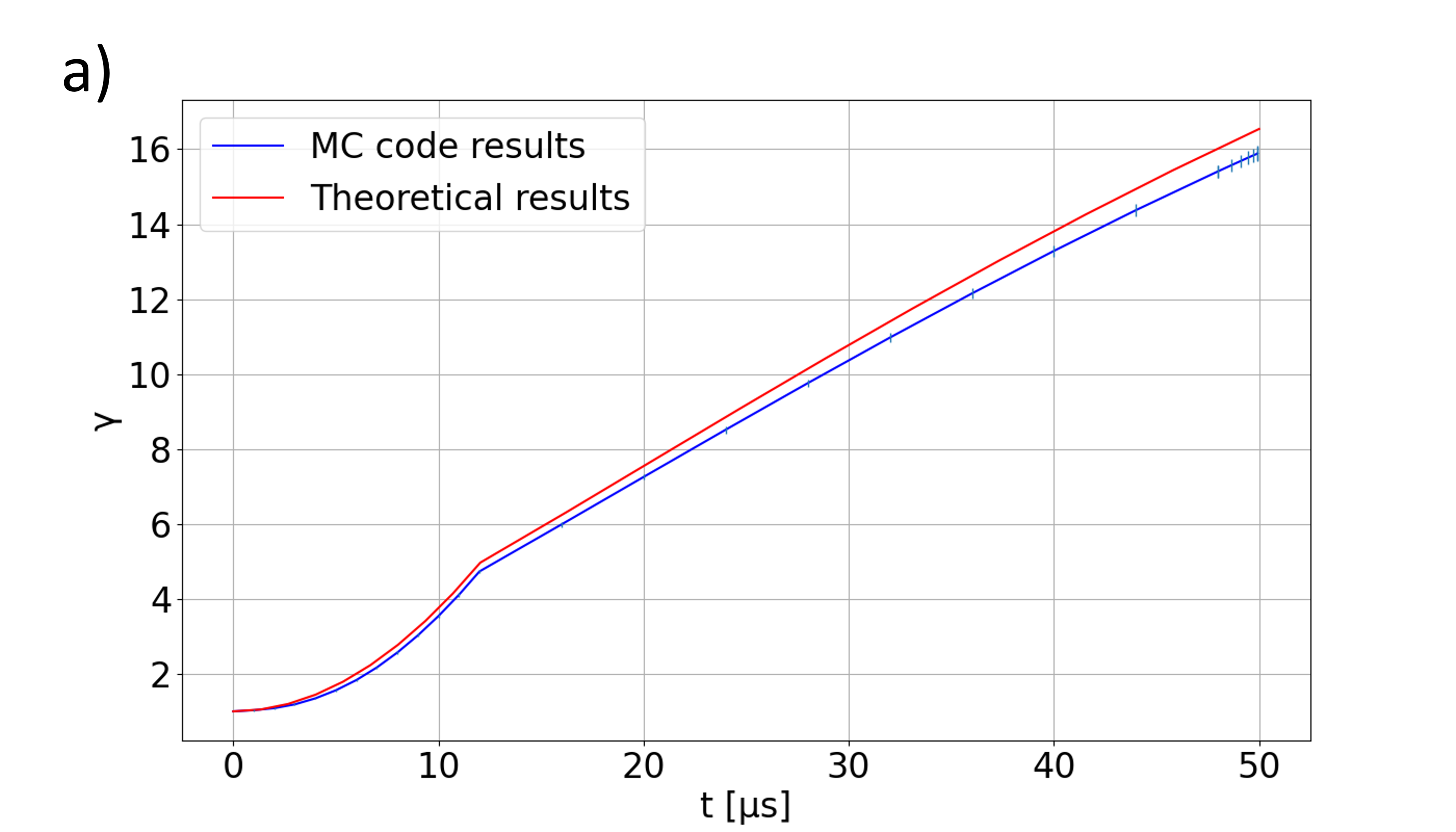}
\end{subfigure}
\begin{subfigure}{1\linewidth}
  \centering
  \includegraphics[width=1\linewidth]{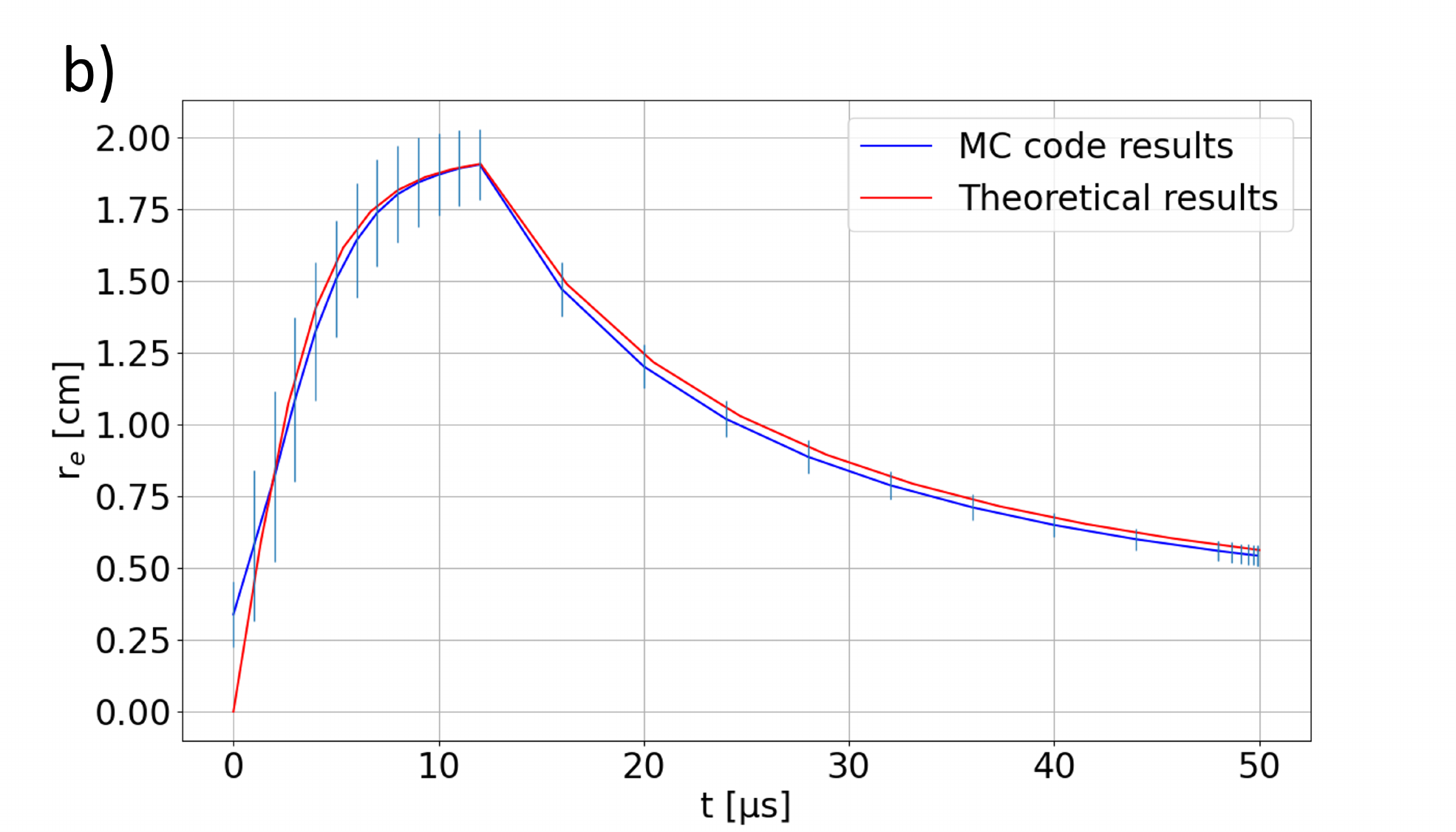}
\end{subfigure}
\caption{Comparison of the temporal evolution during the ECRIPAC working cycle for the He$^{2+}$ prototype design between the average results from the Monte-Carlo (MC) simulations (blue plot) and the theoretical formulation (red plot) for the a) electron Lorentz factor $\gamma$; b) radial distance of the electron from the cavity axis $r_e$.}
\label{fig:mc_comparison}
\end{figure}

\textit{Monte-Carlo code validation --}
A Monte-Carlo particle-tracking code has been used to simulate the electron dynamics, in order to validate the theoretical treatment presented in this paper. The code has already been validated against results from GYRAC accelerators~\cite{Andreev-gyrac2012,Andreev-gyrac2016,Andreev-gyrac2017}. It has been adapted to simulate the prototype He$^{2+}$ accelerator design proposed in the previous paragraph.
The Monte-Carlo code uses a statistic of 100000 electrons, with $\approx$70\% of them undergoing gyromagnetic autoresonance. Further details on the Monte-Carlo code implementation can be found in the supplemental material~\cite{Suppl}. %(see also references \cite{Boris_pusher,Wangler_book,LANL_poisson,Birdsall_book,COMSOL} therein).
Figure~\ref{fig:mc_comparison} compares the average evolution in time of the Lorentz Factor (Fig.~\ref{fig:mc_comparison}a) and the radial distance of the electron from the cavity axis (Fig.~\ref{fig:mc_comparison}b) obtained with the Monte-Carlo code (in blue) and with the theoretical formulation (in red) presented earlier. The results obtained with the Monte-Carlo code show a strong agreement with the theoretical treatment presented in the paper. The energy gain during GA predicted by Eq.~\ref{eq:gyrac_autoresonance} closely follows the average energy of the simulated electrons, while the energy gain during the plasma compression phase (Eq.~\ref{eq:ecripac_compression_gamma}) is slightly overestimated by less than 4\%.
The radial dimension of the electron bunch is also in very close agreement with the theoretical predictions, both during the GA ($r_e=v_e/\Omega$) and plasma compression phase (Eq.~\ref{eq:eq_motion}).
The results of the Monte-Carlo code demonstrate that the theoretical treatment for ECRIPAC can be considered reliable, providing confidence in the study presented in this paper.

% ****** SECTION BREAK ******
\textit{Beam parameters estimation --}
This last section is devoted to the estimation of the beam parameters characterizing the ion bunch extracted from the ECRIPAC accelerator. Once again the compact He$^{2+}$ demonstrator has been considered to provide some numerical results.
The parameters of interest considered in this work are the transverse and longitudinal emittance of the beam, the peak intensity and charge of the extracted ion bunch, the repetition rate, and the required power for the microwave injector, including considerations on the radio-frequency coupling.
Due to the absence of literature on this subject, strong approximations and assumptions are required to provide a tentative estimate for all these quantities. Further details on the calculations presented in this section can be found in the supplemental material~\cite{Suppl}.
\\
The transverse 1-$\sigma$ root mean square (RMS) normalized emittance, taken along the x axis, has been estimated using the usual expression $\epsilon_{x,n}=\gamma\beta \sigma_x \sigma_{x'}$ where $\beta$ is the normalized velocity, $\sigma_x$ is the ion RMS spatial distribution along x and $\sigma_{x'}$ is the RMS of the ion distribution momentum $p_x$ divided by the longitudinal accelerator momentum $p_z$. Assuming ions uniformly distributed inside a disk of radius $r_{e,PL}$ (electron disk radius at the end of the PLEIADE phase), $ \sigma_x  \sim r_{e,PL}/2$.  $\sigma_{x'}$ is estimated by calculating the initial ion transverse momentum spread at the end of the compression phase divided by the final axial ion momentum provided by Eq.~\ref{eq:ecripac-ion_energy_simple}. The obtained value for the transverse emittance is  $\epsilon_{x,n}\approx1.2\cdot10^{-6}$ m$\cdot$rad.
The longitudinal emittance has been estimated using the relation $\epsilon_L=\sigma_E \sigma_t$ where $\sigma_E$ represents the longitudinal bunch energy spread and $\sigma_t$ its temporal width.
Considering Eq.~\ref{eq:pleiade}, we assumed the energy spread of the accelerated ion bunch to be similar to the electron energy spread at the end of the PC phase, which leads to $\sigma_E\approx110$ keV according to the MC simulation results.
Since the axial velocities of the two plasma populations have been experimentally observed to be similar during ion entrainment, the longitudinal size (and hence the bunch temporal width) of both ions and electrons can be considered on the order of the plasma oscillation. Thus, exploiting the MC simulation results, we estimated the bunch temporal width to be $\sigma_t\approx5$ ns.
Hence, the ion bunch longitudinal emittance is estimated to be $\epsilon_{L}\approx4.4\cdot10^{-4}$~eV$\cdot$s.
\\
Regarding the estimation of the ion bunch charge $Q$ and current $I$, considering the ion shake-out effect, an appropriate tuning of the PLEIADE magnetic field should allow the accelerated ion bunch (after electron filtering) to consist solely of the selected charge state. Hence, exploiting the results of the previous calculations, we can estimate that $Q\approx 14$ nC and $I\approx3.3$ A.
\\
In the evaluation of the accelerator repetition rate, the main bottleneck is the frequency of the pulsed coil $f_{pul}$ ruling the GA and PC phases, since the PLEIADE phase is completed on a much shorter timescale (on the order of hundreds of ns). Given the considerable magnetic field intensity that must be generated by the pulsed coil, the repetition rate is mainly limited by the hoop stress ($\sigma_h$) and Joule heating ($W_J$) generated on the magnets during its operation, which both scale as the square of the maximum pulsed field ($\sigma_h,W_J\propto B_{max}^2$) \cite{Herlach-pulsed}. Considering the small dimensions required for the pulsed coil (on the order of some cm), we can extrapolate a realistic repetition rate from pulsed magnets generating a much more intense peak magnetic field~\cite{Yamazaki-pulsed,Linden-pulsed}, obtaining a value in the range of $f_{pul}\approx1-10$ Hz.
\\
Regarding the required RF power for the ECRIPAC, recent experiments on GYRAC accelerators in a long mirror configuration \cite{Andreev-longgyrac2017,Andreev-longgyrac2021}, characterized by the same frequency for the microwave (2.45 GHz) and similar dimensions for the accelerating cavity (radius $R_{cav}=0.5$ cm and length $L_{cav}=80$ cm), used an injected power up to 2.5~kW to successfully achieve gyromagnetic autoresonance.
Given the good agreement between their simulations and experimental results and the similar values of electron density and electric field intensity between their numerical study ($n_e=10^{10}$ cm$^{-3}$ and $E_{HF}=0.3-2.0$ kV/cm) and our MC simulations ($n_e=1.12\cdot10^{10}$ cm$^{-3}$ and $E_{HF}=0.5-2.0$ kV/cm), we can suppose that a similar microwave power will be sufficient for the proposed compact He$^{2+}$ accelerator design.
The RF coupling is extremely difficult to estimate without a full-wave simulation or an experiment. Nevertheless, a large feedback exists on 2.45 GHz ECR ion sources~\cite{Tarvainen-coupling,Jo-coupling} and the GYRAC RF coupling in ECRIPAC is expected to be similar to the one in those plasma sources.

% ****** SECTION BREAK ******

\textit{Conclusion and prospects --}
This work presents a comprehensive introduction to the  ECRIPAC accelerator concept and a summary of its updated theory.
Several accelerator designs with final ion energies from $\approx$~10 to 100 MeV/nucleon, able to generate relevant ion beams for medical application, are presented. A particular focus is devoted to a compact He$^{2+}$ ECRIPAC demonstrator, able to accelerate ion bunches up to 9.5 MeV/A.
A Monte-Carlo Particle-tracking code, already tested on GYRAC accelerator, has been used to validate the theoretical treatment of ECRIPAC using the proposed He$^{2+}$ demonstrator design, showing a very good agreement for the electron dynamics and validating the theoretical study presented in~\cite{Cernuschi-ecripac}.
The ion beam parameters from the detailed He$^{2+}$ compact ECRIPAC accelerator have been estimated using both the MC results and the updated theory. This led to an estimated 1-$\sigma$ RMS normalized transverse emittance $\epsilon_{x,n}\approx1.2\cdot10^{-6}$ m$\cdot$rad, a bunch longitudinal emittance $\epsilon_{L}\approx4.4\cdot10^{-4}$~eV$\cdot$s, a bunch charge and intensity of $Q\approx 14$ nC and $I\approx3.3$ A respectively.
\\
As a prospect, owing to the plasma acceleration nature of ECRIPAC, the use of a self-consistent plasma simulation is deemed necessary to investigate with precision the ion dynamics and the plasma stability in the system. This simulation will also allow to estimate with a much higher accuracy the ion beam parameters extracted from ECRIPAC, to better understand the ion shake-out phenomenon and to further optimize the proposed ECRIPAC designs. Due to the short timescale necessary to resolve gyromagnetic autoresonance, which is considerably smaller than a period of the injected microwave $T_{HF}\approx408$ ps, we assess that a Particle In Cell (PIC) simulation is required, despite its higher computational cost compared to gyro-kinetic or MHD approaches.
A PIC simulation of the entire system during the full working cycle is currently under development. The development time of the future PIC simulation will take full advantage of the present Monte-Carlo study, reducing the computational cost by means of a higher pulsed coil frequency and using the validated demonstrator geometry.

% ****** SECTION BREAK ******

\textit{Acknowledgments --}
We would like to express our sincere gratitude to Patrick Bertrand, a retired theoretician at the GANIL facility, for his valuable insights and extensive information regarding the early investigations of the ECRIPAC accelerator concept.

\bibliography{Bibliography}% Produces the bibliography via BibTeX.

\clearpage
\end{document}